\documentclass[preprint,3p,times,twocolumn,authoryear]{elsarticle}

\usepackage{amssymb}
\usepackage{amsmath}
\usepackage{amsthm}
\usepackage{mathtools}

\usepackage{xcolor}
\newcommand\red[1]{\textcolor{black}{#1}} 

\begin{document}

\begin{frontmatter}



\title{\red{Controlling the snap-through behavior of an elastica strip by cuts}}


\author[inst1,inst2]{Eszter Fehér}

\affiliation[inst1]{organization={Department of Morphology and Geometric Modeling, Faculty of Architecture, Budapest University of Technology and Economics},
            addressline={Műegyetem rkp. 3}, 
            city={H-1111 Budapest},
            country={Hungary}}

\affiliation[inst2]{organization={HUN-REN-BME Morphodynamics Research Group},
            addressline={Műegyetem rkp. 3}, 
            city={H-1111 Budapest},
            country={Hungary}}

\begin{abstract}
This work examines the snap-through behavior of a clamped-clamped \red{elastica} made from initially flat, thin, cut sheets under increasing vertical concentrated load acting at midspan. A two-parameter, symmetric pattern is introduced to conduct a numerical parameter analysis across three different support distances. When the support distance is one-quarter of the total length of the sheet, the structure loses stability at a symmetry point bifurcation over a wide range of parameters. Additionally, there exists a small range of parameters where limit point bifurcation occurs. In this case, the cuts can induce symmetry in the stability loss. For larger support distances (half or three-quarters of the total length), limit point bifurcation occurs only for small cuts, and there is a range of cut parameters that leads to monotonic monostability, indicating that no stability loss occurs. These findings are supported by experimental data. Overall, our research demonstrates that carefully designed cut patterns can either control the mode of stability loss in \red{elastica strips} or suppress it entirely.

\end{abstract}

%


\end{frontmatter}



\section{Introduction}
\label{sec:introduction}
\red{Tailoring the geometry and mechanical properties of structures composed of thin sheets can be achieved by cuts and perforation. \citet{Shrimali2021} investigated how the perforation of a plate can alter its bending properties. Similarly, cuts made on inextensible sheets can induce various mechanical properties, such as stretchability, auxetic behavior, and programmable kinematics \citep{Tao2023}. Structures assembled from cut sheets have a wide range of engineering applications, including ventilation systems \citep{Montalvo2024}, space structures \citep{Pedivellano2024}, evaporator systems \citep{Li2024}, biomedical devices \citep{Wu2025}, soft robotics \citep{Hong2022}, electronics \citep{Liu2022}, and metamaterials \citep{Hamzehei2024}. Many of them are inspired by the paper technique known as kirigami, that enables the creation of three-dimensional structures from flat sheets through cutting and folding. Since these structures are made from flat sheets, they are material-efficient, easy to transport, and ideal for deployable designs.}

\red{These structures can lose their stability while still remaining functional or even gain new functionalities. It is possible to design the mode of stability loss and the behavior that follows \citep{Rafsanjani2017, Du2024}. This design approach allows dome-shaped perforated shells to switch between multiple stable states through a snap-through mechanism \citep{Yang2024, Cho2023}. Many kirigami structures consist of bending-active strips created by bringing the ends together to form an elastica shape \citep{Lee2022,Hong2022}, that can snap-through under external loads. By perforating the sheet or introducing additional cuts, we can alter the geometry \citep{Zhang2022,Liu2020,Ye2024} or the structural rigidity of an individual strip \citep{Feher2024,Feher2026b}. However, there is still a lack of literature exploring the relationship between cut patterns and the stability properties of perforated strips.}

The literature on the stability of arches is extensive. \citet{Timoshenko1961} focused on determining the critical buckling load of curved, circular arches. \citet{Pi1998} and later \citet{Pi2002} analyzed the stability behavior shallow circular arches. \citet{Luu2016} explored elliptical arches with various support conditions and identified different types of buckling, including cases where no buckling occurred. Their findings indicated that non-slender structures and shallow arches did not experience buckling.

There are also several studies related to the stability properties of the elastica.  \citet{Chen2011} conducted a snap-through analysis of an elastica subjected to a concentrated vertical load, where the supports are hinges. The analysis revealed that different load conditions lead to varying stability properties, such as sub-critical pitchfork bifurcation (symmetry point bifurcation) or limit point bifurcation. Additionally, \citet{Zhang2025} explored the snap-through behavior of a heavy, hard magnetic elastica. \citet{Gomez2017} investigated the dynamics of a snapping elastica, while \citet{Liu2016} focused on a constrained elastica designed as an energy harvester. Furthermore, \citet{Cheng2023} examined the buckling behavior of compressed, constrained ribbons under vertical loading, and \citet{Zhang2020} investigated the buckling modes of buckled ribbons when subjected to vertical loads.
\red{Changing the boundary conditions could also lead to snap-through. \citet{Cazzolli2019} determined the range of boundary angles corresponding to snapping (snap surface) by the analytical solution of the Elastica equation. \citet{Cazzolli2020} introduced an elastica catastrophe machine that allows one end of the elastica to move manually. They determined a closed region in which the end can be freely moved without snapping while crossing the region boundary results in snapping. Here, we also aim to determine regions of different stability behaviors by analyzing the effect of cuts.}

\red{None of these works on the elastica problem considered the effects of cuts introduced in the geometry. Introducing cuts on slender structures could effect their stability properties. An example of the power of cutting is demonstrated by \citet{Fauli2026}. They showed that introducing slits on compressed beams could be used to control their postbuckling behavior.}

In this study, we investigate the snap-through instability of a \red{partially cut elastica} depending on various cut patterns. Our analysis focuses on a clamped-clamped \red{elastica} made from an initially flat, thin, rectangular sheet, which is subjected to a vertical concentrated load at its midpoint. We introduce a simple, two-parameter, symmetric cut pattern consisting of two rectangular cuts placed along the longer edges of the sheet. Our objective is to show that the cut pattern can control the stability behavior of the arch. To achieve this, we conduct a numerical parameter analysis and compare our findings with experimental results. 

The paper is structured as follows: Section \ref{sec:Methodology} outlines the mechanical model and numerical approach, Section \ref{sec:results} presents the numerical and experimental results along with their discussion, and Section \ref{sec:conclusion} concludes the paper.

\section{Methodology}
\label{sec:Methodology}

\subsection{Mechanical background}
\begin{figure}[t]
    \centering
    \includegraphics[keepaspectratio,width=0.9\linewidth]{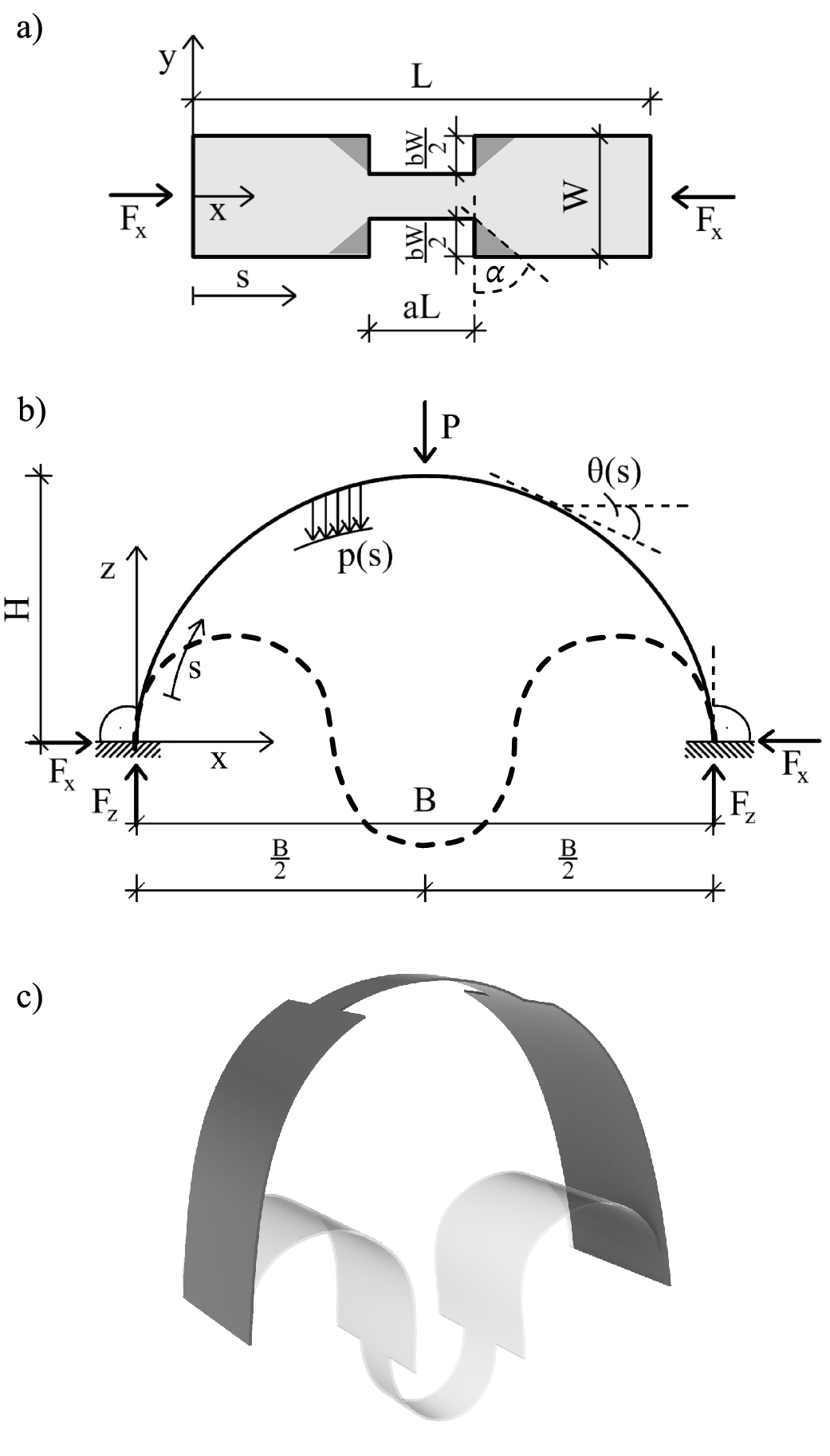}
    \caption {a) Illustration of the cut pattern and its \red{dimensionless} parameters $a$ and $b$. Dark grey regions are considered to be stress-free, and they are neglected in the calculation of the effective moment of inertia of the sheet. b) Mechanical model of the structure. The dashed thick line is the inverted shape after snap-through. c) Three-dimensional visualization of the structure in the initial state ($P=0$) and after snap-through ($P>P_{crit}$).}
    \label{fig:model}
\end{figure}

We consider a thin, initially flat rectangular sheet with thickness $t$, length $L$, and width $W$. The sheet has two transversally symmetric rectangular holes positioned at the center of its longer edges, characterized by \red{dimensionless} parameters $a$ and $b$ (Fig. \ref{fig:model}a). The sheet is forced to take an arched shape by bringing its ends closer together at a distance of $B<L$ and fixing them at vertical positions (Fig. \ref{fig:model}b). Two horizontal forces $F_x$ are applied at the ends of the sheet to maintain equilibrium in the arched form. Additionally, the sheet carries its self-weight $p$, along with a concentrated load $P$ acting at the midpoint. If all the parameters are fixed, the cut pattern determines the shape of the arch. In general, as the concentrated load $P$ is increased, the structural height decreases, and the shape loses stability at $P_{\mathrm{crit}}$. If the supports are not too close to each other, the structure takes an inverted shape (Fig. \ref{fig:model}b,c). This phenomenon is called the snap-through. 

\begin{figure}[t]
    \centering
    \includegraphics[keepaspectratio,width=0.7\linewidth]{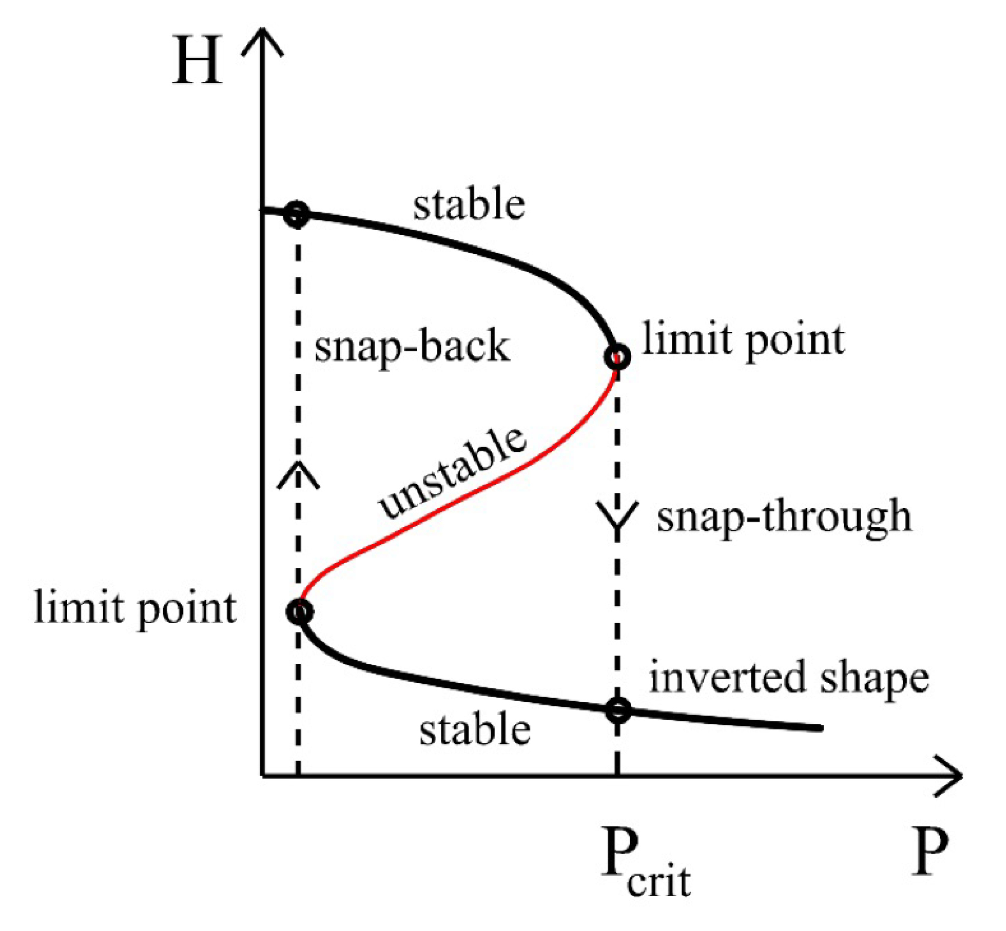}
    \caption {Schematic load-height diagram visualizing a limit point bifurcation. Red and black denote the unstable and stable points, respectively.}
    \label{fig:PH_diagram}
\end{figure}

We assume that some parts of the sheet near the cuts \red{are inactive, i.e., they} do not contribute to carrying the loads. These regions form isosceles right triangles ($\alpha = \pi/4$) that are removed from the cross-section (Fig. \ref{fig:model}a). \red{As a result, the small triangles near the cuts contribute only to the self-weight and their effect on the shape is assumed to be negligible, leading to a continuously varying, symmetric cross-section that can be modeled as a beam. Since the load acts at the center, the cut patterns is transversally symmetric and stability loss is initiated by the deformations of the midline, there is no need for more advanced two-dimensional models. However, more complicated cut patterns with closely placed, asymmetric, internal cuts might need plate models to account for two-dimensional effects.}

We consider quasi-static loading, examining both stable and unstable equilibrium states, and do not consider the dynamic aspects of snap-through. Due to the small thickness and transverse symmetry of the sheet, we can use a nonlinear beam model, the elastica equation, to describe the deformations of the sheet. The sheet is parameterized by its arc-length $s \in [0,L]$ and the deformations are described by the tangent angle $\theta (s)$. The concentrated load acts at $s_P=L/2$. The moment equilibrium equation \citep{howell2009applied} is:
\begin{equation} \begin{split} \label{eq:elastica}
E\frac{\mathrm{d}}{\mathrm{d}s} \left[ I_{\mathrm{eff}}(s) \frac{\mathrm{d}\theta(s)}{\mathrm{d}s} \right] - F_z \cos{\theta(s)} + F_x \sin{\theta(s)} \\ + \cos{\theta(s)} \int_{0}^s Qp(s) \,\mathrm{d}s + P \cos{\theta(s)} \hat{H}(s-s_P) = 0,
\end{split}
\end{equation}
where $E$ is the Young's modulus; $I_{\mathrm{eff}}(s)$ is the effective moment of inertia; $\hat{H}$ is the Heavyside function; $p(s)$ is the self-weight distribution, normalized such that $\int_0^Lp(s)ds = 1$ and scaled by the load intensity $Q$; $P$ is the concentrated vertical load; $F_x$ and $F_z$ are the horizontal and vertical support reactions, respectively. 

\red{The effective width of the sheet $w_{\mathrm{eff}}(s)$ is the remaining width after the removal of the inactive parts:
\begin{equation}\label{eq:weff}
w_{\mathrm{eff}}(s) = \begin{cases*}
  \min(W,(1-b)W-2(s-s_l)\tan\beta), & if $s <= s_l$,\\
  (1-b)W,              				& if $s_l < s < s_r$,\\
  \min(W,(1-b)W+2(s-s_r)\tan\beta), & if $s >= s_r$,
\end{cases*}
\end{equation}
where $s_l = \frac{(1-a)L}{2}$, $s_r = \frac{(1+a)L}{2}$ and $\beta = \pi/2-\alpha$. 
The corresponding effective moment of inertia is:
\begin{equation}\label{eq:Ieff}
I_{\mathrm{eff}}(s) = w_{\mathrm{eff}}(s)t^3/12. 
\end{equation}}

There are four boundary conditions
\begin{equation} \label{eq:bc1}
\theta(0) = \pi/2,
\end{equation}
\begin{equation} \label{eq:bc2}
\theta(L) = - \pi/2,
\end{equation}
\begin{equation} \label{eq:bc3}
x(L) = \int_0^{L} \cos(\theta(s)) \, \mathrm{ds} = B,
\end{equation}
\begin{equation} \label{eq:bc4}
z(L) = \int_0^{L} \sin(\theta(s)) \, \mathrm{ds} = 0
\end{equation}
prescribing the tangent and the position of the sheet at the supports, respectively.

The structural height $H$ is the vertical position of the midpoint, i.e., $H=z(L/2)$. The load-height diagram of the system is shown in Fig. \ref{fig:PH_diagram}. During unloading, the load gradually decreases, and the inverted shape loses stability at a limit point, resulting in the so-called snap-back. An unstable set of points connects the upper and lower stable parts of the diagram. 

Depending on the distance between the supports, even though the setup is symmetric, the stability loss might be possible before the limit point. In this case, the symmetric solution becomes unstable, and the structure passes through non-symmetric shapes before settling at a lower stable position \citep{Thompson1973}, a phenomenon called symmetric point bifurcation. 

It is important to note that our problem has some physical constraints that are not incorporated in the model. The beam model (Eq. \ref{eq:elastica}) permits self‐intersections, and the supports restrict only the positions of the endpoints (Eqs. \ref{eq:bc1}–\ref{eq:bc4}), without imposing any limitations on the internal points. Consequently, points of the arch may pass through the supports in the model. This additional freedom requires extra caution when comparing the results with experimental observations.

\subsection{Numerical procedure}

Equation \eqref{eq:elastica} has no analytical solution in the general case. To compute both stable and unstable equilibrium configurations, we use a combination of Chebyshev polynomial approximation and numerical continuation, a methodology previously successfully applied by \citet{Feher2026a} to determine equilibrium shapes of a curved rod under distributed loading. The structure is discretized by $n$ points, and the derivatives of $\theta$ are approximated by derivatives of Chebyshev polynomials using Chebfun \citep{driscoll2014chebfun} in Matlab. 

The Chebyshev approximation converges exponentially to the target function as $n$ increases, even for the derivatives \citep{Trefethen2000}. \red{Although working with Chebyshev polynomials requires dense matrix operations, the number of mesh points can be kept relatively low. However, using low number of mesh points can cause sampling errors. Moreover, the cut pattern defined by Eq. \eqref{eq:weff} is piecewise linear that could lead to approximation errors when using Chebfun such as the Gibbs phenomenon \citep{driscoll2014chebfun}. Here, we prevented the occurence of these errors by applying a small rounding off at the internal corners of the pattern.}

\begin{figure}[htbp]
    \centering
    \includegraphics[keepaspectratio,width=\linewidth]{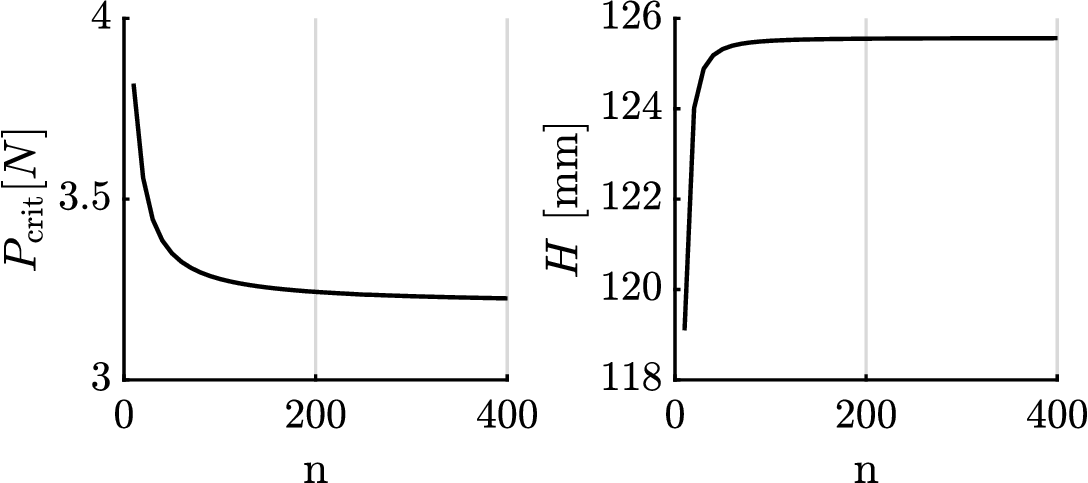}
    \caption {\red{Convergence of the critical load (left) and the initial height ($P=0$) (right) in terms of the mesh size for a full sheet ($a=b=0$) with $B=0.5L$.}}
    \label{fig:full_convergence}
\end{figure}

\begin{figure}[htbp]
    \centering
    \includegraphics[keepaspectratio,width=\linewidth]{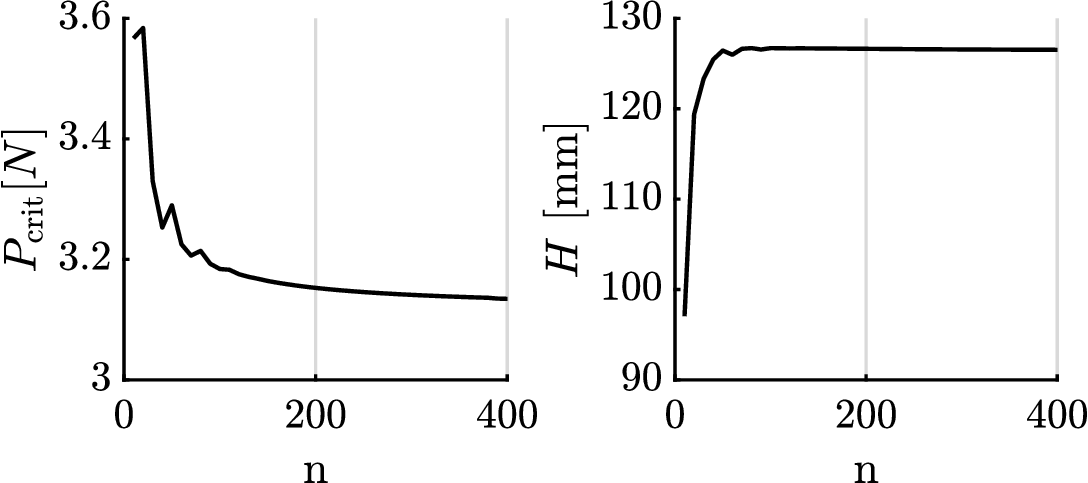}
    \caption {\red{Convergence of the critical load (left) and the initial height ($P=0$) (right) in terms of the mesh size for a cut sheet ($a=0.2, b = 0.1$) with $B=0.5L$.}}
    \label{fig:cut_convergence}
\end{figure}

 \red{We carried out convergence tests for the full sheet (Fig. \ref{fig:full_convergence}) and the cut sheets  (Fig. \ref{fig:cut_convergence}) to determine the mesh size and selected $n=340$, which provides sufficient resolution even in the vicinity of the cuts.} The boundary conditions given by Eqs. \eqref{eq:bc1} and \eqref{eq:bc2} were applied directly in the $n$ resulting equations, and we added two additional equations for Eqs. \eqref{eq:bc3} and \eqref{eq:bc4}. This leads to a total of $n+2$ equations. 

The resulting system of equations was solved via numerical continuation using pde2path \citep{dohnal2014pde2path, uecker2021numerical}, treating $F_x$ and $F_z$ as free parameters. First, we start from a stable initial configuration with $P=Q=0$, and perform continuation in $B$ until the target value is achieved. Subsequently, $B$ is fixed at its target level, and the continuation proceeds in $Q$. In the final step, we take $P$ as the continuation parameter and track the eigenvalues of the Jacobian of the system. Eigenvalues with negative real parts indicate unstable solutions \citep{uecker2021numerical}. A bifurcation point occurs when an eigenvalue crosses zero, and its location is determined using bisection. The tolerance for the residual is set to $10^{-8}$, and the maximum step size was $1/20$ of the target parameter value. 

Using this procedure, we calculated the $H-P$ diagrams for various support distances, as well as different $a$ and $b$ values, and determined the corresponding critical loads. 

\section{Results and discussion}
\label{sec:results}
\subsection{Parameter study} \label{sec:parameter_study}

\begin{figure*}[htbp]
    \centering
    \includegraphics[keepaspectratio,width=\linewidth]{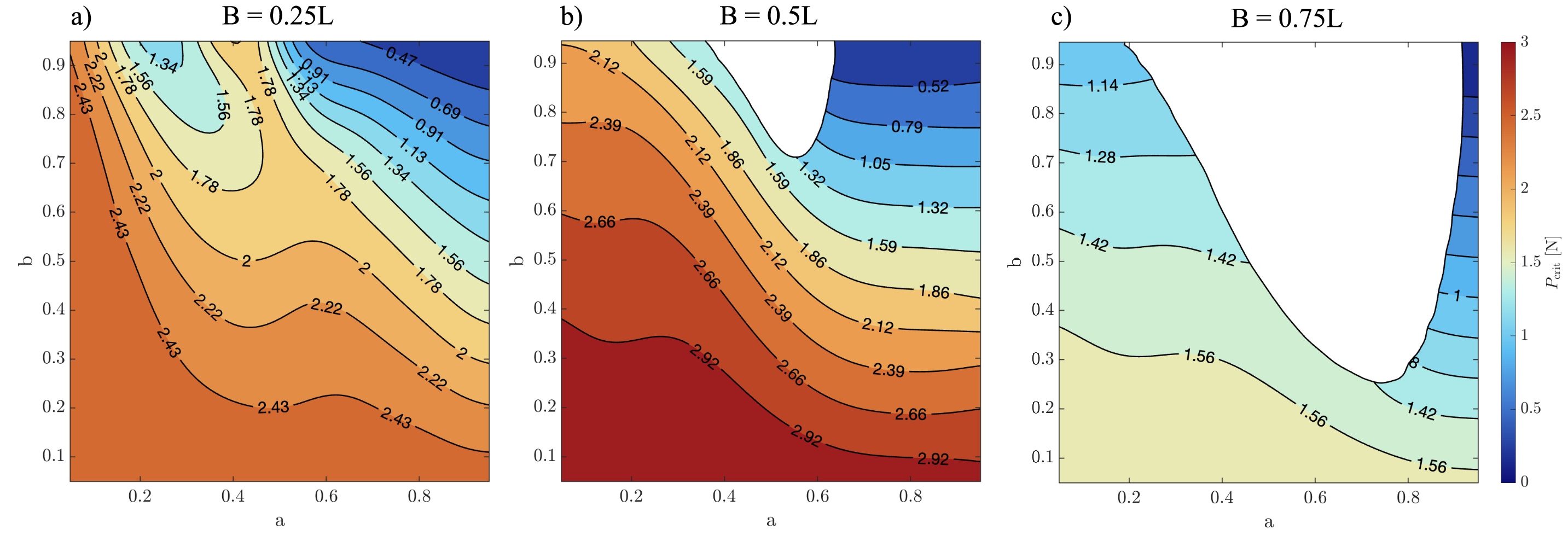}
    \caption {Critical value of the load in terms of the \red{dimensionless} pattern parameters $a$ and $b$. a) For $B=0.25L$, all the patterns lead to stability loss. b) In case $B=0.5L$, there is no stability loss in the white region. c) In case $B=0.75L$, there is no stability loss in the white region.}
    \label{fig:Pcrit_ab}
\end{figure*}

\begin{figure*}[htbp]
    \centering
    \includegraphics[keepaspectratio,width=\linewidth]{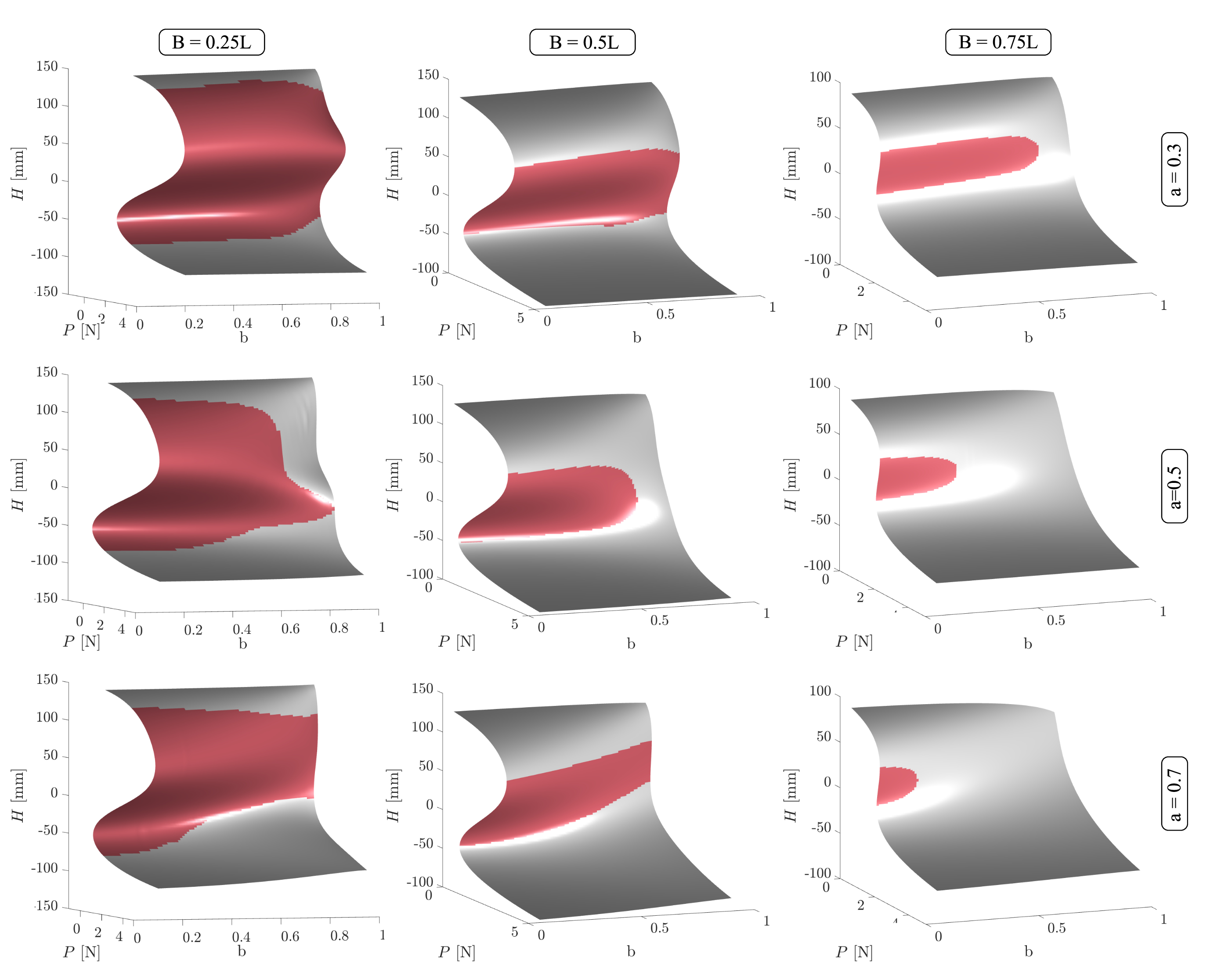}
    \caption {Shaded illustration of equilibrium surfaces for different $a$ parameters (rows) and $B$ support distances (columns). Grey and red colors correspond to stable and unstable states, respectively.}
    \label{fig:branches}
\end{figure*}

\begin{figure*}[htbp]
    \centering
    \includegraphics[keepaspectratio,width=\linewidth]{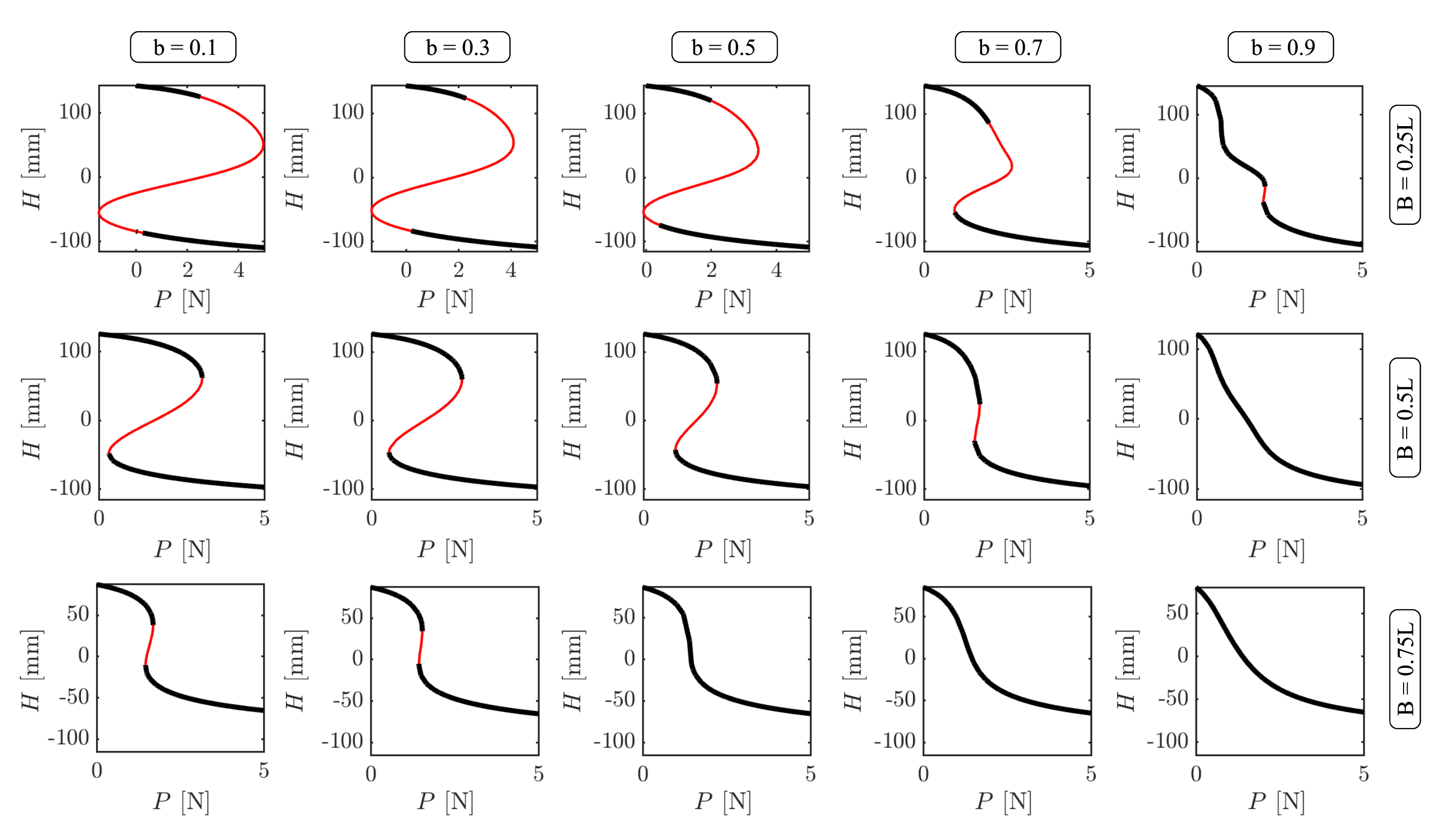}
    \caption {$P-H$ diagrams for different support distances (rows). The length of the cut, $a=0.5$, was kept constant, and only the width, $b$, changed (columns). Black and red correspond to stable and unstable configurations, respectively.}
    \label{fig:single_branches}
\end{figure*}

We generated cut patterns with $0.05\leq a \leq0.95$ and $0.05\leq b \leq0.95$ and computed the load-height diagram for three different support distances: $B=0.25L$, $0.5L$, and $0.75L$. The resulting critical loads in terms of the pattern parameters are shown in Fig. \ref{fig:Pcrit_ab}. Interestingly, some patterns exhibited no snap-through at all for $B=0.5$ and $B=0.75L$, indicating monotonic monostability. These cases correspond to the white regions in Fig. \ref{fig:Pcrit_ab}b-c. 

As expected, smaller cuts generally lead to higher critical loads, and the largest cut ($a=b=0.95$) resulted in the lowest critical load for all $B$ values. However, the decrease rate of $P_{crit}$ is not uniform across the parameter range. This behavior can be linked to the resulting structural shapes. Small $a$ and large $b$ values are cuts that behave similarly to hinges in the structure. Previous work \citep{Feher2026b} showed, that certain cut configurations can lead to structurally preferable shapes under a given load. Thus, while the smallest cut indeed yields the highest critical load, the favorable geometric effect of hinge-like cuts mitigates the weakening expected for patterns with small $a$ values. 

Figure \ref{fig:branches} represents the $P-b-H$ equilibrium surfaces for all investigated support distances for three representative cut lengths $a=0.3,0.5,0.7$. Stable and unstable states are marked in grey and red, respectively. The upper boundary of the red region corresponds to the $P_{crit}$ values. 

Sections of the equilibrium surfaces corresponding to $a=0.5$ (middle row of Fig. \ref{fig:branches}) are presented in Fig. \ref{fig:single_branches} for $b=0.1,0.3,0.5,0.7$, and $0.9$.

\subsection{Limit point bifurcation ($B=0.5L$, $B=0.75L$)} \label{sec:lim_bif}
For $B=0.5L, a=0.5,0.7$ and $B=0.75L,  a=0.3,0.5$, we observe that the unstable region in Fig. \ref{fig:branches} diminishes as $b$ increases. In these cases, the equilibrium surfaces progressively flatten out, and the limit points vanish as $b$ becomes large (Fig. \ref{fig:single_branches}). The surface does not flatten out in the case of $B=0.5L$, $a=0.3,0.7$ since the $a=0.5$ vertical line does not cross the white region in Fig. \ref{fig:Pcrit_ab}b. Monotonic monostability occurs only for a certain range of parameters. 

Previous studies have shown that monotonic monostability can also be achieved in circular curved beams by reducing slenderness or height \citep{Luu2016}. A possible explanation for the transition from snap-through monostability to monotonic monostability, as seen in Figs. \ref{fig:branches}-\ref{fig:single_branches}, lies in the change in the structural shape. Local weakening alters the unloaded shape, which can in turn modify the stability characteristics. Softened regions have lower bending resistance, allowing the structure to deform smoothly without reaching a limit point. Moreover, snap-through requires an initially stiff response. Softening the structure reduces the initial resistance, flattening the force–height curve and causing the limit point to disappear.

\subsection{Symmetry point bifurcation ($B=0.25L$)} \label{sec:symm_bif}

\begin{figure}[htbp]
    \centering
    \includegraphics[keepaspectratio,width=\linewidth]{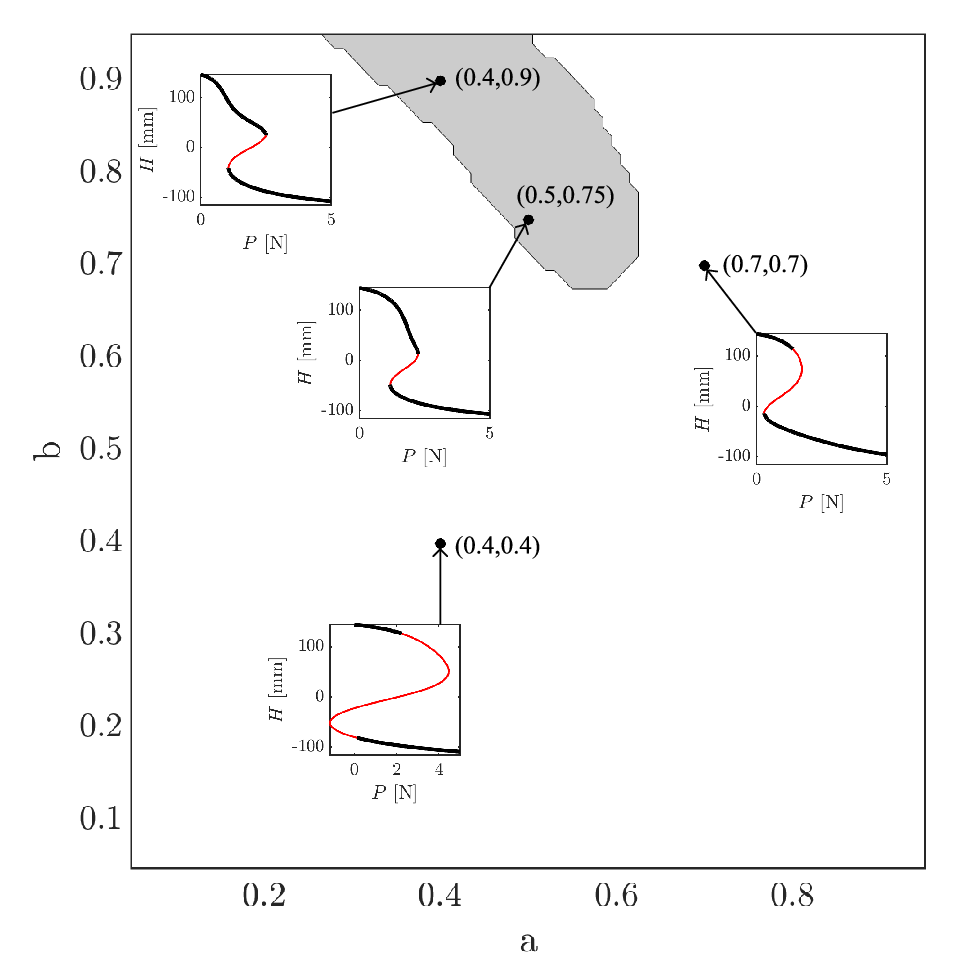}
    \caption {Grey zone corresponds to $a$ and $b$ parameters that exhibit limit point bifurcation, while the structure goes through symmetry point bifurcation outside the grey zone. The $P-H$ branch for some representative parameter combinations is shown in the inset figures, where red and black indicate unstable and stable configurations, respectively.}
    
    \label{fig:asym}
\end{figure}

\begin{figure}[t]
    \centering
    \includegraphics[keepaspectratio,width=\linewidth]{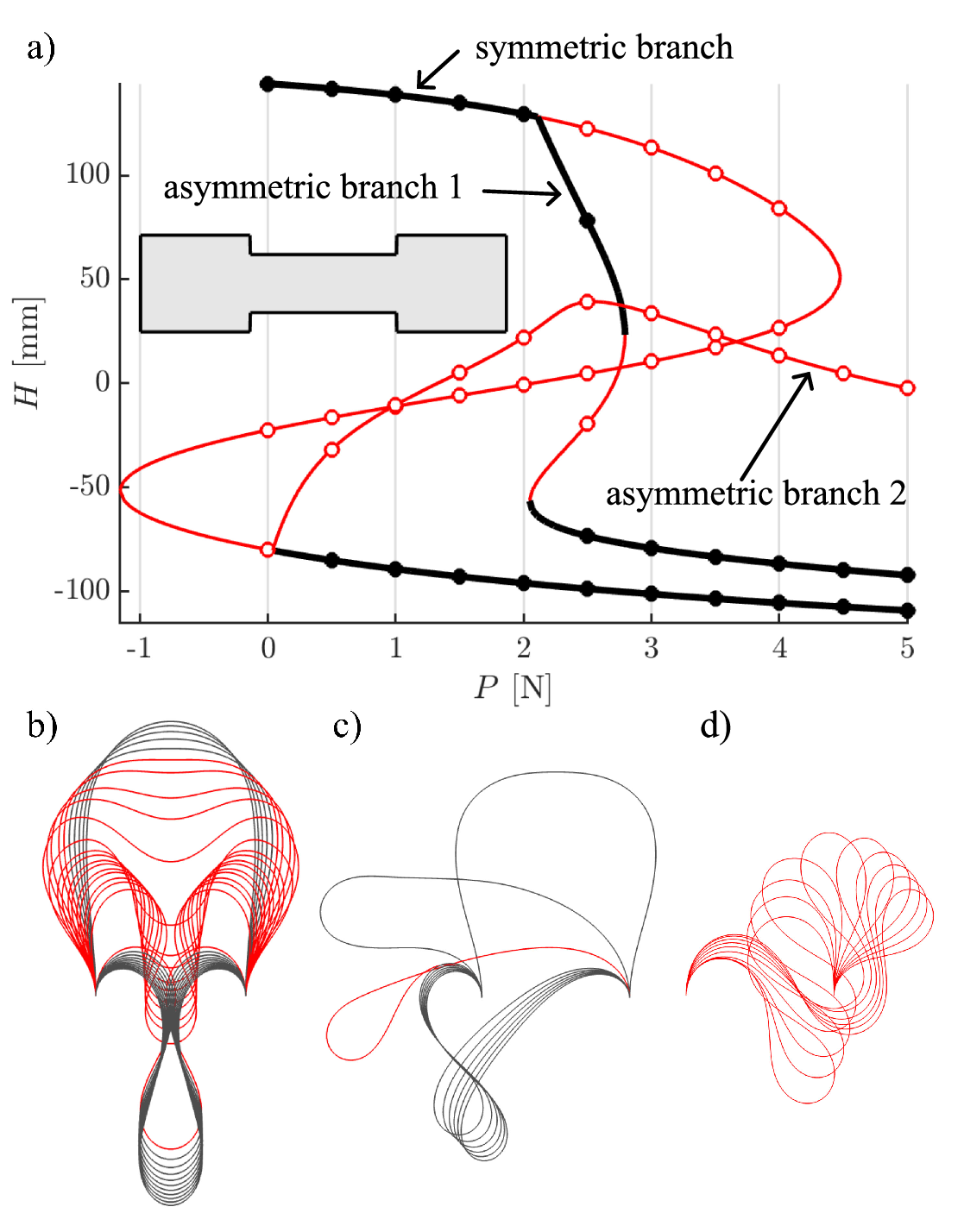}
    \caption {Numerical results of the pattern $a=0.4$, $b=0.4$. a) Symmetric and asymmetric branches of the $P-H$ diagram. Black and red color corresponds to stable and unstable states, respectively. The inset visualizes the pattern. b) Stable and unstable shapes (black and red, respectively) corresponding to the marked points on the symmetric branch. \red{c)-d) Stable and unstable shapes (black and red, respectively) corresponding to the marked points on the asymmetric branches 1 and 2, respectively.}}
    \label{fig:a04b04}
\end{figure}

In this subsection, we take a closer look at the case of $B=0.25L$, that significantly differs from the other two analyzed support distances. 

Figure \ref{fig:branches} shows that the unstable region starts before the fold points for nearly all values of $a$ and $b$. For $a=0.5$, increasing $b$ causes the surface to flatten out at the top, and a new limit point emerges in the middle, where the stability loss coincides with a limit point. For $a=0.7$, although the surface also tend to flatten, the unstable region does not decrease significantly. 

Consequently, not only symmetry point bifurcation but also limit point bifurcation can occur for this support distance. These two modes can be distinguished numerically by checking whether the onset of negative eigenvalues of the Jacobian coincides with the limit points, i.e., the location of the extreme values of the $H-P$ function. Figure \ref{fig:asym} shows the mode of bifurcation in the $a-b$ parameter space, where grey and white regions correspond to limit point and symmetry point bifurcations, respectively. A set of representative load-height diagrams is shown as inset figures.

The $P-H$ diagrams we presented so far correspond only to the symmetric branch. However, the numerical tool we use allows branch switching at bifurcation points. Fig. \ref{fig:a04b04}a presents both the symmetric and asymmetric branches for $a=0.4, b=0.4$. \red{As the load increases, there is a supercritical pitchfork bifurcation, where asymmetric shapes become stable and the symmetric shape becomes unstable. The asymmetric shapes loose stability at a limit point. Following the diagram from the bottom direction, i.e., decreasing the load, there is a subcritical pitchfork bifurcation, where both the asymmetric and symmetric shapes become unstable. The corresponding structural shapes for the symmetric and asymmetric branches are shown in Fig. \ref{fig:a04b04}b and c-d, respectively. Note that there are two branches overlapping on Fig. \ref{fig:a04b04}a for each asymmetric branches corresponding shapes in Fig. \ref{fig:a04b04}c-d and their reflection.}

These findings suggest that, in a load-controlled experiment with sufficiently small load increments, the structure would naturally transition onto the asymmetric branch. The lower stable portions of each path are physically inadmissible because the structure intersects itself. Determining the physically realizable configurations would require solving the constrained elastica problem, which lies beyond the scope of this work.

Enforcing symmetric snap-through rather than asymmetric buckling is achieved by softening the middle region of the structure. It allows the arch to deflect more under the same load, bringing it closer to the inverted configuration and enforcing symmetric stability loss. Without sufficient cuts, the structure undergoes asymmetric buckling. There are two competing effects: while cuts soften the structure, they also reduce the critical load. Consequently, only a limited range of cut sizes can induce symmetric snap-through. If the cuts are too large, the critical load associated with the limit-point bifurcation becomes so low that the softening effect is suppressed.

\subsection{Experiments} \label{sec:experiments}

\begin{figure*}[htbp]
    \centering
    \includegraphics[keepaspectratio,width=\linewidth]{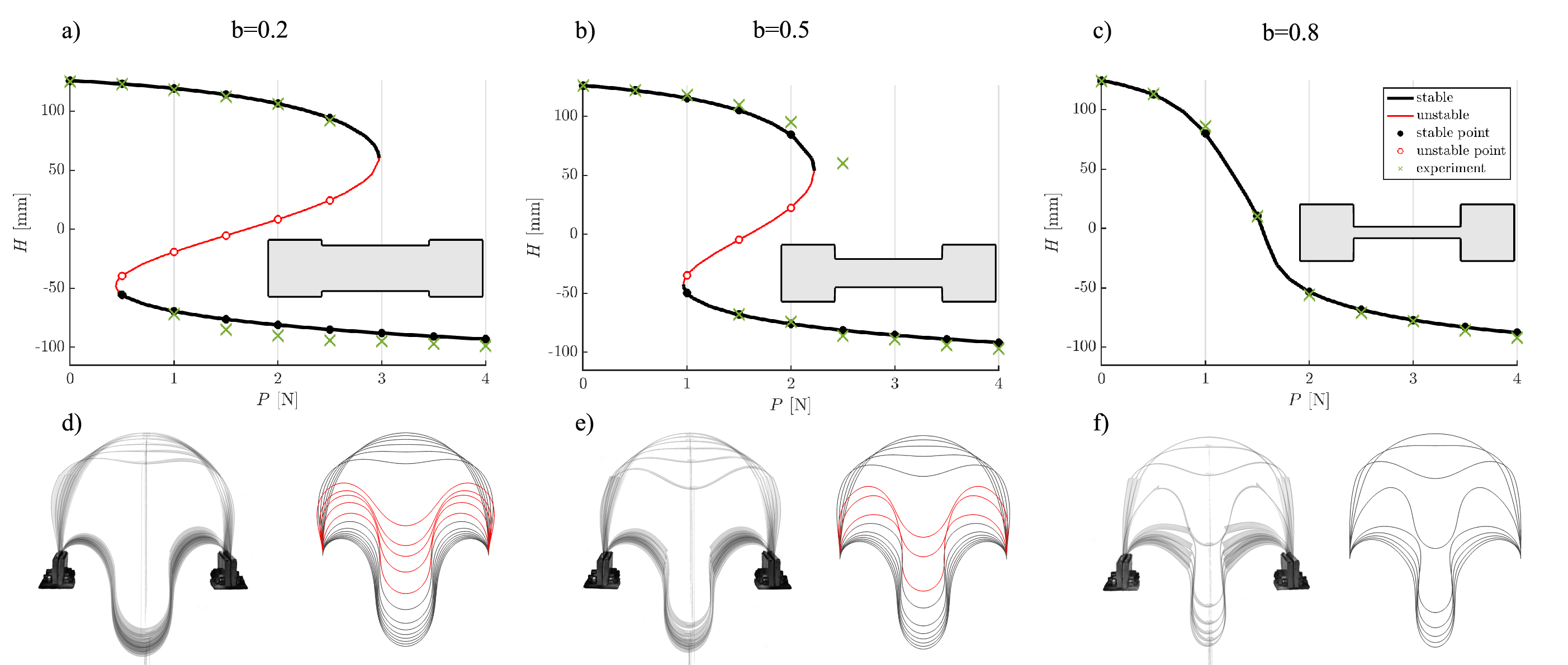}
    \caption {Experimental and numerical results for $B=0.5L$. The length of the cut was kept $a=0.5$. a)-c) Force-height diagrams for different cut heights. Black and red lines correspond to stable and unstable points, respectively. The markers are placed in $0.5\, N$ steps. Black and red circular markers correspond to stable and unstable points, respectively. Green 'x' denotes the measured height values. The inset illustrates the cut pattern. a) $b=0.2$, b) $b=0.5$, c) $b=0.8$. d)-f) Photographs (left) and calculated shapes (right)  corresponding to the markers. Red shows unstable shapes. d) $b=0.2$, e) $b=0.5$, f) $b=0.8$.}
    \label{fig:exp05}
\end{figure*}

The experiments were performed on laser-cut PET sheets with thickness $t=0.5\,\mathrm{mm}$. The Young's modulus of the sheet is $E=2800\,\mathrm{N/mm^2}$ \citep{polymer}, and the density $\rho=1.33\,\mathrm{g/cm^3}$ was specified by the manufacturer. Three cut patterns were selected, all with fixed cut length $a=0.5$ and varying cut heights $b=0.2,0.5,0.8$. Each pattern was tested for three support distances $B=0.25L,0.5L$, and $0.75L$. The bounding rectangle of the patterns were 360x90 mm, with $10\,\mathrm{mm}$ used for the clamps at each end, leaving $L=340\,\mathrm{mm}$, $W=90\,\mathrm{mm}$. A small hole of $1\,\mathrm{mm}$ radius was cut at the center of the sheet to accommodate the load, which was neglected in the calculations. 

The load was applied on the arches in $0.5\,\mathrm{N}$ steps. The height of the structure was measured at the center of the sheet, using a measuring bar placed next to the structure. \red{The measurement uncertainty of the height was $\pm$0.5 mm.} We photographed the structure and measured the height at each load step. After reaching the highest load level, the structure was unloaded in $0.5\,\mathrm{N}$ steps to capture the lower stable branch and the snap-back behavior. 

\begin{figure*}[htbp]
    \centering
    \includegraphics[keepaspectratio,width=\linewidth]{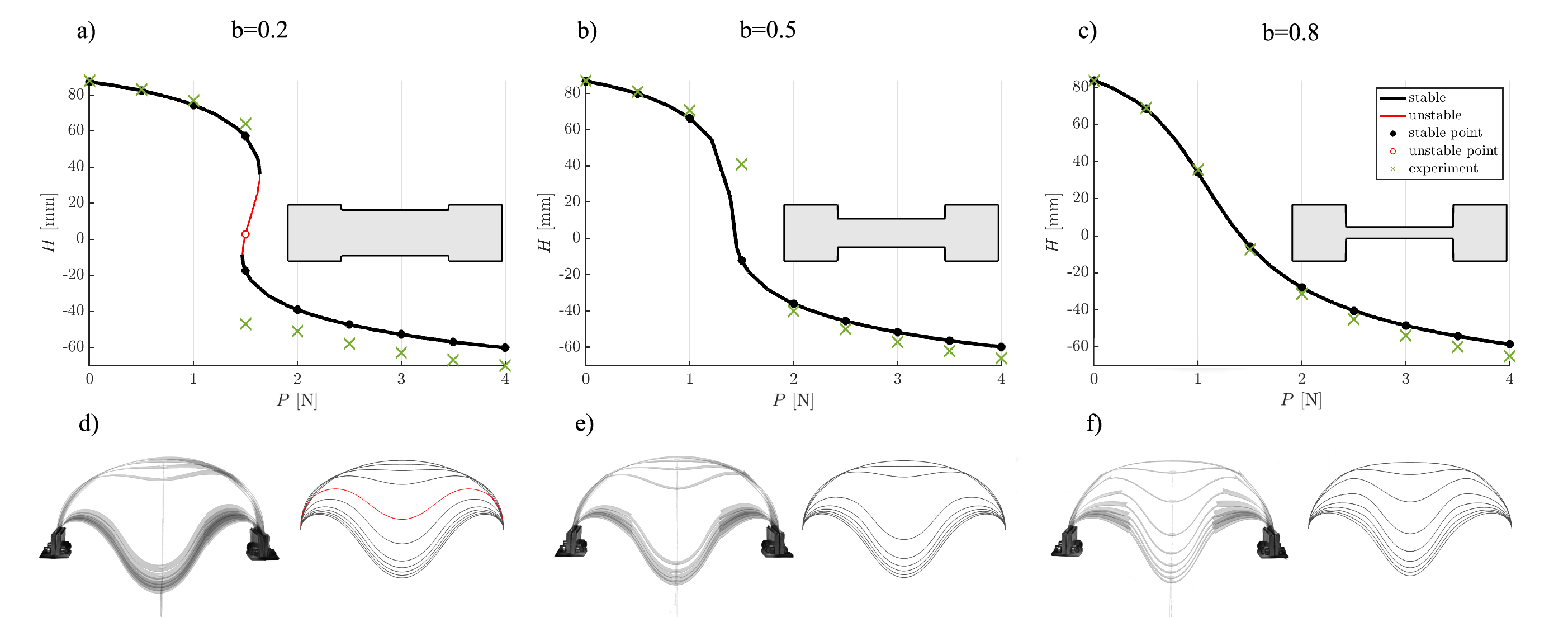}
    \caption {Experimental and numerical results for $B=0.75L$. The length of the cut was kept $a=0.5$. a)-c) Force-height diagrams for different cut heights. Black and red lines correspond to stable and unstable points, respectively. The markers are placed in 0.5 steps. Black and red circular markers correspond to stable and unstable points, respectively. Green 'x' denotes the measured height values. The inset illustrates the cut pattern. a) $b=0.2$, b) $b=0.5$, c) $b=0.8$. d)-f) Photographs (left) and calculated shapes (right) corresponding to the markers. Red shows unstable shapes. d) $b=0.2$, e) $b=0.5$, f) $b=0.8$.}
    \label{fig:exp075}
\end{figure*}

We begin by discussing the cases exhibiting a limit-point bifurcation. Fig. \ref{fig:exp05} compares the numerical predictions and experimental results for $B=0.5L$. The measured heights are plotted along with the calculated $P-H$ diagram in Fig. \ref{fig:exp05}a-c. Overall, there is a good agreement between the calculations and the experimental data. The largest differences occur near the bifurcation points, likely due to their sensitivity to small inaccuracies in the cut pattern, the applied load, or the load position. The calculated structural shapes also closely match the photographs (Fig. \ref{fig:exp05}d-f), and the structure remained symmetric at each load level. For $b=0.8$ (Fig. \ref{fig:exp05}c), no snap-through or snap-back was observed, i.e., the loading and unloading paths coincided. This behavior is consistent with the numerical predictions, which indicate monotonic monostability for this pattern.

\begin{figure*}[htbp]
    \centering
    \includegraphics[keepaspectratio,width=\linewidth]{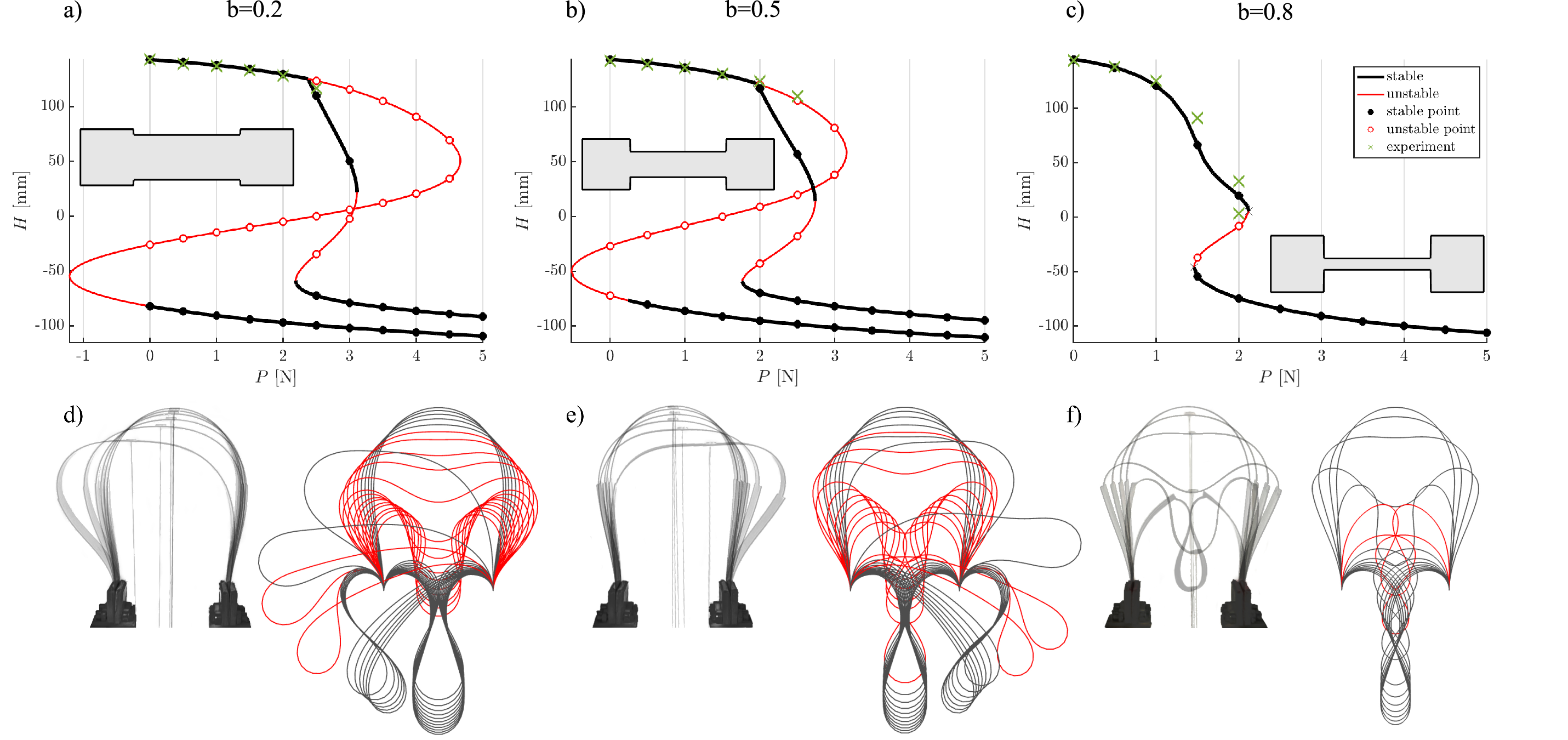}
    \caption {Experimental and numerical results for $B=0.25L$. The length of the cut was kept $a=0.5$. a)-c) Force-height diagrams for different cut heights. Black and red lines correspond to stable and unstable points, respectively. The markers are placed in 0.5 steps. Black and red circular markers correspond to stable and unstable points, respectively. Green 'x' denotes the measured height values. The inset illustrates the cut pattern. a) $b=0.2$, b) $b=0.5$, c) $b=0.8$. d)-f) Photographs (left) and calculated shapes (right) corresponding to the markers. Red shows unstable shapes. d) $b=0.2$, e) $b=0.5$, f) $b=0.8$. }
    \label{fig:exp025}
\end{figure*}

The case of $B=0.75L$ also showed good agreement between numerical predictions and experimental results (Fig. \ref{fig:exp075}). Only $b=0.2$ exhibited snap-through, with the largest differences between the calculations and experiments occurring near the bifurcation point (Fig. \ref{fig:exp075} a). For $b=0.5$, no snap-through was observed; however, the middle part of the diagram is very steep, which corresponds to the region of largest deviation between experiments and calculations. Regions with gentler slopes exhibited better agreement. The closest agreement was achieved for $b=0.8$ (Fig. \ref{fig:exp075}c), where the $P-H$ diagram is free from steep parts and stability loss. In the experiments, both $b=0.5$ and $b=0.8$ showed no snap-through or snap-back, in excellent agreement with the numerical predictions of monotonic monostability. 

As discussed in Subsection \ref{sec:symm_bif}, the case of $B=0.25L$ is more complex. The selected $b$ values include both limit point and symmetry point bifurcations. Additionally, at high load levels, some solutions become physically unfeasible due to self-intersections. Our experimental setup also imposed a constraint on the horizontal displacement of the arch, preventing testing at the highest load levels. 

Figure \ref{fig:exp025}a,b shows the load-height diagrams for $b=0.2$ and $b=0.5$, including both symmetric and asymmetric branches. \red{As for the latter, we plotted only asymmetric branch 1, corresponding to the supercritial bifurcation point.} The corresponding experimental and calculated shapes are shown in Fig. \ref{fig:exp025}d,e. All calculated symmetric and asymmetric shapes are overlaid for comparison. The photographs reveal that the structure initially follows the symmetric branch; once the symmetric shape loses stability, it transitions to an asymmetric configuration. Further loading was limited by the horizontal displacement constraint in the experiments. 

The lower stable branches of the $P-H$ diagrams in Fig. \ref{fig:exp025} contain self-intersections or contacting parts. Consequently, we could compare only the upper parts of the diagrams to experiments. The experiments and calculations agreed well, the largest difference occurred near the bifurcation points.

Figure \ref{fig:exp025}c corresponds to a case with limit point bifurcation, where the structure remained symmetric in both the experiments and the calculations. The calculated and measured heights agree well, and the largest differences occur near the steep parts of the diagram. The structural shapes also match closely (Fig. \ref{fig:exp025}f). The photographs show that the last shape (corresponding to $P=2\,\mathrm{N}$) has contacting parts, indicating that the assumptions of the model no longer hold beyond this point.

\section{Conclusion}
\label{sec:conclusion}
The results reveal an interesting phenomenon: by carefully designing the cut patterns, it is possible to control the mode of stability loss \red{for a partially cut elastica} under vertical loads. We demonstrated examples in which snap-through disappears for certain cut patterns, as well as cases in which cuts enforce symmetric stability loss. As a result, cuts can significantly alter the overall stability characteristics.

For larger support distances ($B=0.5L$ and $B=0.75L$), the structure goes through snap-through at a limit point bifurcation in general, but the limit point disappears for certain $a$ and $b$ parameter pairs and the structure goes to the inverted position through stable shapes without exhibiting snap-through. Therefore, introducing cuts can suppress stability loss and enforce monotonic monostability. 

For $B=0.25L$, the symmetric shape loses stability and the physically observable equilibrium path is the asymmetric path for a wide range of parameters. However, there is a small range of cut patterns that keeps the symmetric path stable until the limit point, leading to a limit point bifurcation. As a result, cuts can enforce symmetric stability loss. 

In summary, we identified patterns with limit point bifurcation, symmetry point bifurcation, and monotonic monostability. The phenomena observed here arise from multiple competing effects, including local softening, and structural shape. 

Overall, our results highlight the potential of \red{cutting} to tailor the stability behavior of \red{bending-active structures made of thin sheets.} In addition to the behaviors studied here, we could expect bistable, multistable, or more complex responses for more intricate cut patterns. The results can be applied to adjustable energy absorbers, deployable structures, and energy harvesting systems. \red{Future work could explore the simultaneous optimization of geometry and mechanical response to achieve desired stability characteristics.} 

\section*{Acknowledgments}
We acknowledge the Digital Government Development and Project Management Ltd. for awarding us access to the Komondor HPC facility based in Hungary. This research was supported by the NKFIH Hungarian Research Fund Grants 143175. This research was conducted with the support of the ERASMUS+ Programme of the European Union.


\bibliographystyle{elsarticle-harv} 
\bibliography{cas-refs}


%
%
%
%
\end{document}